\documentclass[twocolumn,aps,prl,showpacs,superscriptaddress]{revtex4}
\usepackage{amssymb}
\usepackage{graphicx}

\begin{document}
\title{Experimental Long-Distance Decoy-State Quantum Key Distribution Based On Polarization Encoding}
\author{Cheng-Zhi Peng}
\thanks{These authors contributed equally to this work.}
\affiliation{Department of Physics, Tsinghua University, Beijing 100084, China}
\affiliation{Hefei National Laboratory for Physical
Sciences at Microscale and Department of Modern Physics,
University of Science and Technology of China, Hefei, Anhui
230026, China}
\author{Jun Zhang}
\thanks{These authors contributed equally to this work.}
\affiliation{Hefei National Laboratory for Physical
Sciences at Microscale and Department of Modern Physics,
University of Science and Technology of China, Hefei, Anhui
230026, China}
\author{Dong Yang}
\thanks{These authors contributed equally to this work.}
\affiliation{Department of Physics, Tsinghua University, Beijing 100084, China}
\author{Wei-Bo Gao}
\affiliation{Hefei National Laboratory for Physical
Sciences at Microscale and Department of Modern Physics,
University of Science and Technology of China, Hefei, Anhui
230026, China}
\author{Huai-Xin Ma}
\affiliation{Department of Physics, Tsinghua University, Beijing 100084, China}
\affiliation{China Electronics System Engineering Company, Beijing 100039, China}
\author{Hao Yin}
\affiliation{China Electronics System Engineering Company, Beijing 100039, China}
\author{He-Ping Zeng}
\affiliation{Key Laboratory of Optical and Magnetic Resonance Spectroscopy and
Department of Physics, East China Normal University, Shanghai 200062, China}
\author{Tao Yang}
\affiliation{Hefei National Laboratory for Physical Sciences at Microscale and Department
of Modern Physics, University of Science and Technology of China, Hefei, Anhui
230026, China}
\author{Xiang-Bin Wang}
\affiliation{Department of Physics, Tsinghua University, Beijing 100084, China}
\author{Jian-Wei Pan}
\affiliation{Department of Physics, Tsinghua University, Beijing 100084, China}
\affiliation{Hefei National Laboratory for Physical Sciences at Microscale and Department
of Modern Physics, University of Science and Technology of China, Hefei, Anhui
230026, China}
\affiliation{Physikalisches Institut, Universit\"at Heidelberg, Philosophenweg 12, 69120 Heidelberg, Germany}
\date{\today}

\begin{abstract}
We demonstrate the decoy-state quantum key distribution
(QKD) with one-way quantum communication in polarization
space over 102~km. Further, we simplify the experimental
setup and use only one detector to implement the one-way
decoy-state QKD over 75~km, with the advantage to overcome
the security loopholes due to the efficiency mismatch of
detectors. Our experimental implementation
can really offer the unconditionally secure final keys. We
use 3 different intensities of 0, 0.2 and 0.6 for the
light sources in our experiment. In order to eliminate
the influences of polarization mode dispersion in the
long-distance single-mode optical fiber, an automatic
polarization compensation system is utilized to implement
the active compensation.
\end{abstract}


\maketitle


Quantum key distribution~\cite{BB84,GRTZ02,DLH06} can in
principle offer the unconditionally secure private
communications between two remote parties, Alice and Bob.
However, the security proofs for the ideal BB84
protocol~\cite{M01,SP00} do not guarantee the security of a
specific setup in practice due to various imperfections
there. One important problem in practical QKD is the
effects of the imperfect source, say, the coherent states.
The decoy state method~\cite{H03, Wang05,
Wang05_2,LMC05,HQph} or some other methods~\cite{SARG04,
K04} can help to generate the unconditionally secure final keys even
an imperfect source is used by Alice in practical QKD.
Basically, QKD can be realized in both free space and
optical fiber~\cite{GRTZ02}. Each option has its own
advantages. The fiber QKD can be run in the always-on mode:
it runs in both day and night and is not affected by the
weather. Also, the future local QKD networks are supposed
to be using fiber. So far, there are many experiments of
fiber QKD with weak coherent lights~\cite{QKD}. However,
these results actually do not offer the unconditional
security because of the possible photon-number-splitting
attack~\cite{PNS}. Recently, there are also experimental
implementations of the decoy-state method~\cite{Lo06}, with two-way
quantum communication. However, since these implementations
have not taken the specific operations as requested by
Ref.~\cite{GFKZR05}, the security of the final keys is still
unclear due to the so-called Trojan horse
attacks~\cite{GRTZ02}. One can implement active
counter-measures~\cite{GFKZR05} to overcome this problem,
which is deserved experimental implementations in the future.
The other way is to use one-way quantum communication which we
have adopted in this work.

Here we present the first polarization-based decoy-state QKD
implementation over 102~km with only one-way quantum communication
using two detectors and 75~km using only one detector.
Our results are unconditionally secure (For the unconditional
security, we mean that the probability that Eve has non-negligible
amount of information about the final key is exponentially close
to 0, say, $e^{-O(100)}$). Here we must clarify that given the
existing technologies~\cite{QKD}, if the distance is shorter than
about 20~km, through the simple worst-case
estimation~\cite{GLLP04} of the fraction of tagged bits it is
still possible for one to implement the unconditionally secure QKD
without using the decoy-state method.

We can know how to distill the secure final keys with imperfect
source given the separate theoretical results from
Ref.~\cite{GLLP04}, if we know the upper bound of the fraction of
tagged bits (those raw bits generated by multi-photon pulses from
Alice) or equivalently, the lower bound of the fraction of untagged
bits (those raw bits generated by single-photon pulses from Alice).
In Wang's 3-intensity decoy-state protocol~\cite{Wang05,Wang05_2},
one can randomly use 3 different intensities (average photon
numbers) of each pulses (0, $\mu$, $\mu'$) and then
observe the counting rates (the counting probability of Bob's
detector whenever Alice sends out a state) of pulses of each intensities
($S_0, S_{\mu},S_{\mu'}$). The density operators for the states of
$\mu$ and $\mu'$ ($\mu'>\mu$) are
\begin{eqnarray}
\left\{\begin{array}{ll}
\rho_\mu=e^{-\mu}|0\rangle\langle0|+\mu e^{-\mu}|1\rangle\langle 1|+c\rho_{c} \\
\rho_{\mu'}=e^{-\mu'}|0\rangle\langle0|+
\mu'e^{-\mu'}|1\rangle\langle 1|+\frac{\mu'^{2}e^{-\mu'}}{\mu^{2}e^{-\mu}}c\rho_{c}+d\rho_d,
\end{array} \right.
\label{eqs}
\end{eqnarray}
here $c={1-e^{-\mu}-\mu e^{-\mu}}$,$\rho_c=\frac{e^{-\mu}}{c}\sum_{n=2}^{\infty}\frac{\mu^n}{n!}|n\rangle\langle n|$,
$\rho_d$ is a density operator and $d>0$ (here we use the same notations in Ref.~\cite{Wang05,Wang05_2}). We denote $s_0(s_0'),s_1(s_1'),s_c(s_c')$ for the
counting rates of those
vacuum pulses, single-photon pulses and $\rho_c$ pulses from
$\rho_\mu(\rho_{\mu'})$.
Asymptotically, the values of primed symbols here should be equal to
those values of unprimed symbols. However, in an experiment the
number of samples is finite therefore they could be a bit different.
The bound values of $s_1,s_1'$ can be determined by the following
joint constraints corresponding to Eq.(15) of Ref.~\cite{Wang05_2}
\begin{eqnarray}\label{num}
\left\{\begin{array}{ll}
S_\mu=e^{-\mu}s_{0}+\mu e^{-\mu}s_{1}+cs_{c} \\
cs_{c}' \leq \frac{\mu^{2}e^{-\mu}}{\mu'^{2}e^{-\mu'}}(S_{\mu'}-
\mu'e^{-\mu'}s_{1}'-e^{-\mu'}s_{0}'),
\end{array} \right.
\end{eqnarray}
where $s_{1}'=(1-\frac{10e^{\mu / 2}}{\sqrt{\mu s_{1}N_{\mu}}})s_{1}$,
$s_{c}'=(1-\frac{10}{\sqrt{s_{c}N_{\mu}}})s_{c}$, $s_{0}'=0$,
$s_{0}=(1+r_0)S_0$ and $r_0=\frac{10}{\sqrt{S_{0}N_{0}}}$ to obtain the worst-case
results~\cite{Wang05,Wang05_2}. $N_\mu,N_0$ are the pulse numbers
of intensity $\mu,0$. Given these, one can calculate
$s_1,s_{1}',s_c$ numerically.


The experimental setup is shown in Fig.1, mainly including transmitter
(Alice), quantum channel, receiver (Bob) and electronics system. All
the electronics modules are designed by ourselves. The synchrodyne
(SD) is designed by field programmable gate array (FPGA, Altera Co.)
and outputs multiple channels of synchronous clocks with independent
programmable parameter settings, which is equivalent to an arbitrary
function generator, to drive the modules of random number generator
(RNG), data acquisition (DAQ, designed by FPGA) and single-photon detector (SPD)
respectively. The signals with FWHM of about 1~ns are generated by laser diode
driver (LDD) to drive 10 distributed feed-back laser diodes (LD) at the central
wavelength of 1550~nm, where 4 LDs are used for decoy states ($\mu$)
and another 4 LDs are used for signal states ($\mu'$) and the other
2 LDs are used for polarization calibration. The polarization states
of photons emitting from LDs can be transformed to arbitrary
polarization state by polarization controller (PC). For decoy states
and signal states, the four polarization states are $|H\rangle,
|V\rangle, |+\rangle, |-\rangle$, where $|H\rangle, |V\rangle$
represent horizontal polarization and vertical polarization,
$|+\rangle = 1/\sqrt{2}(|H\rangle + |V\rangle)$ and $|-\rangle =
1/\sqrt{2}(|H\rangle - |V\rangle)$, as the four states for the
standard BB84 protocol~\cite{BB84}. For test states, the two
polarization states are $|H\rangle$ and $|+\rangle$ to calibrate the
two sets of polarization basis. The photons of
every channel are coupled to an optical fiber via fiber coupling
network (FCN), which is composed of multiple beam splitters (BS) and
polarization beam splitters (PBS) and optical attenuators. In FCN,
the fiber length of every channel must be adjusted precisely so that
the arrival time differences to SPD caused by the fiber length
differences can be less than 100~ps.

In the setup, a dense wavelength division multiplexing fiber filter (FF) is inserted
 in Alice's side. On the one hand, it can guarantee that
the wavelengths of emitted photons in all channels are equal to
avoid the possibility of Eve's attack utilizing the variance
of photon wavelengths. On the other hand, it can reduce the
influences of chromatic dispersion in fiber.

\begin{figure}
\centerline{\includegraphics[scale=0.98]{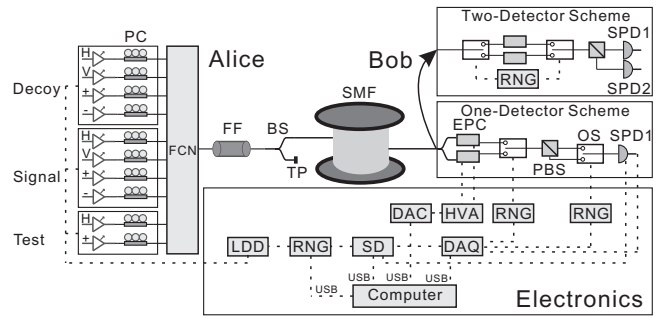}}
\caption{\label{Fig1} Schematic diagram of the experimental
setup. Solid line and dashed line represent optical fiber
and electric cable respectively. See the text for
the abbreviations.}
\end{figure}

In the experiment, the pulse numbers ratio
of the 3 intensities is $5:4:1$
and the intensities of signal states and decoy states
are fixed at $\mu' = 0.6$ and $\mu = 0.2$ respectively, which
are not necessarily the optimized parameters with our setup. The
fluctuations of intensities are monitored at the test point (TP).
If the varying range of fluctuation is larger than about 5\%
we will stop the system and adjust the light sources.
In fact, the effects of intensity fluctuation is indeed
a very important theoretical problem for the decoy-state QKD, which had not been
solved by theorists prior to our experiment.
Very recently theoretical progress has shown that the effects
are moderate with certain modifications of the experimental setup~\cite{Wang06}.

After passing through the long-distance single-mode fiber
(SMF, Corning Co., fiber attenuation is about 0.2~dB/km),
at Bob's side we adopt two kinds of
measurement scheme. In one-detector scheme, firstly a
fiber BS is used to select the two polarization
basis called $HV$ basis and $+-$ basis randomly. Secondly due to the
polarization mode dispersion (PMD) effects in long-distance
SMF, we develop an automatic polarization compensation
(APC) system to compensate for the PMD actively. The principles
of APC are: Alice sends fixed $|H\rangle$ states or
$|+\rangle$ states. Then Bob records the accepted counting
rates in the corresponding basis using DAQ system and transmits them to the
computer via universal serial bus (USB). After algorithmic
processing, the computer gives out the data, which can be
converted to voltages of electric polarization controllers
(EPC, General Photonics Co.) through digital-to-analog
converter (DAC) and high voltage amplifier (HVA). Then the
fiber squeezers in EPC are driven by the voltages and
change the polarization~\cite{EPC}. After repeating
feedback controls the visibility of test states
can reach the target value and the APC system stops.
The average adjusting time is about 3 minutes.
However, this time can be greatly shortened
using a continuous coherent light and wavelength division multiplexing techniques and optimized algorithms.
Figure 2(a) shows the test results of
the APC system with 75~km fiber. In the experiment,
Alice starts APC to calibrate the system first. After calibration
she transmits the pulses for several minutes. Then this process is
repeated. Thirdly we use two magnetic optical switch (OS, Primanex Co.) with
switching time of less than 20~$\mu s$ driven
by two independent RNG to randomly switch the
basis and the output ports of PBS respectively. The one-detector scheme
can overcome the security loopholes due to the efficiency mismatch of
detectors~\cite{TSA} since the fiber lengths of each state can be adjusted identically.
On the other hand, two OS driven by
the same RNG and two SPDs~\cite{Note1} are used in two-detector scheme to
reduce the transmission losses in Bob's side.

\begin{figure}
\centerline{\includegraphics[scale=0.80]{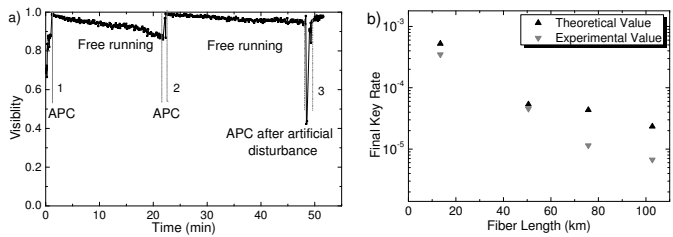}}
\caption{\label{Fig2} a) Test of the APC system with 75~km fiber.
In position 1 and 2 the APC system monitors the visibility
changes of polarization states and adjusts the voltages of
EPC actively to reach the target visibility. Subsequently,
the visibility of polarization states becomes worse slowly
when free running. In position 3, artificial disturbance
induces drastic change of visibility and the APC system
can still work well. The time interval of points is 2 seconds.
b) Comparison of the final key
rate of signal states per pulse between the
theoretical calculation and experimental results with
four different distance settings $L$
(13.448~km, 50.524~km, 75.774~km and 102.714~km), where
their corresponding total attenuations are
(24.9~dB, 32.2~dB, 34.8~dB, 37.0~dB) including channel
losses, all the insertion losses of components,
detector efficiencies etc.
The first three settings use one-detector scheme and
the last one uses two-detector scheme. The first two
settings' repetition frequency is $f$=4~MHz and the last
two is $f$=2.5~MHz.}
\end{figure}


We use this all-fiber quantum cryptosystem to implement the
decoy-state QKD over 102~km and 75~km using two-detector scheme
and one-detector scheme respectively. The experimental parameters and their
corresponding values are listed in Table~\ref{tab:table1}. In the
experiment, Alice totally transmits about $N$ pulses to Bob. After
the transmission Bob announces the pulse sequence numbers and
basis information of received states. Then Alice broadcasts to Bob
the actual state class information
and basis information of the corresponding pulses.
Alice and Bob can calculate the
experimentally observed quantum bit error rate (QBER) values
$E_{\mu}, E_{\mu'}$ of decoy states and signal states
according to all the decoy bits and a small fraction
of the signal bits respectively~\cite{Note2}.


\begin{table}
\caption{\label{tab:table1}Experimental parameters (P.) and their
corresponding values of 75.774~km (Value1) and 102.714~km (Value2) decoy-state QKD.}
\begin{ruledtabular}
\begin{tabular}{llllll}
P. & Value1 & Value2 & P. & Value1 & Value2\\
\hline
$L$ & 75.774~km & 102.714~km & $S_{\mu'}$ & $2.076\times 10^{-4}$ & $1.262\times 10^{-4}$\\
$f$ & 2.5~MHz & 2.5~MHz & $S_{\mu}$ & $7.534\times 10^{-5}$ & $4.611\times 10^{-5}$\\
$N$ & $1.607\times 10^{9}$ & $5.222\times 10^{9}$ & $S_{0}$ & $9.174\times 10^{-6}$ & $6.711\times 10^{-6}$\\
$E_{\mu'}$ & 3.231\% & 3.580\% & $s_{1}'$ & $2.460\times 10^{-4}$ & $1.558\times 10^{-4}$\\
$E_{\mu}$ & 9.039\% & 9.098\% & $R_{\mu'}$ & $1.143\times 10^{-5}$ & $6.706\times 10^{-6}$\\
$E_{1}^{\mu'}$ & 6.099\% & 5.854\% & $R_{E}$ & 11.668~Hz & 8.090~Hz\\
$R_{T}$ & 46.167~Hz & 29.427~Hz & $\frac{R_{E}}{R_{T}}$ & 0.253 & 0.275\\
\end{tabular}
\end{ruledtabular}
\end{table}

Then we can numerically calculate a tight lower bound of the
counting rate of single-photon $s_{1}'$ using Eq.~\ref{num}.
The next step is to estimate the fraction of single-photon $\Delta_{1}$
and the QBER upper bound of single-photon $E_{1}$. We use
\begin{equation}
\label{Del}
\Delta_{1}^{\mu'}=s_{1}'\mu'e^{-\mu'}/S_{\mu'}, \Delta_{1}^{\mu}=s_{1}\mu e^{-\mu}/S_{\mu}
\end{equation}
to conservatively calculate $\Delta_{1}$ of signal states and
decoy states respectively~\cite{Wang05,Wang05_2}.
And $E_{1}$ of signal states and decoy states
can be estimated by the following formula,
\begin{equation}
\label{Eq}
E_{1}^{\mu'(\mu)}=(E_{\mu'(\mu)}-\frac{(1-r_0)S_{0}e^{-\mu'(\mu)}}{2S_{\mu'(\mu)}})/\Delta_{1}^{\mu'(\mu)}.
\end{equation}
Here we consider the statistical fluctuations of the
vacuum states to obtain the worst-case results.

Lastly we can calculate the final key rates of signal states
using the formula~\cite{Wang05,Wang05_2} of
\begin{equation}
\label{Eq2}
R_{\mu'}=S_{\mu'}(\Delta_{1}^{\mu'}-H(E_{\mu'})-\Delta_{1}^{\mu'}H(E_{1}^{\mu'})),
\end{equation}
here $H(x)=-xlog_{2}(x)-(1-x)log_{2}(1-x)$.
Then we compare the experimental final key rate of signal states
$R_{E}$ with the theoretically allowed value $R_{T}$, i.e., in
the case both $\Delta_{1}$ and $E_1$ are known without any overestimation. The
theoretically allowed values of $\Delta_{1}$ and $E_1$ for
signal states are
\begin{eqnarray}
\left\{\begin{array}{ll}
\Delta_{1T}^{\mu'}=(S_{\mu'}-(1-\mu')S_{0})e^{-\mu'}/S_{\mu'}\\
E_{1T}^{\mu'}=(E_{\mu'}-\frac{S_{0}e^{-\mu'}}{2S_{\mu'}})/\Delta_{1T}^{\mu'},
\end{array} \right.
\label{eq:The}
\end{eqnarray}
with the assumptions that the ideal value of the single-photon
counting rate is $s_{1T}=\eta+S_{0}$ and $S_{\mu'}=\eta\mu'+S_{0}$,
where $\eta$ is the overall transmittance. We find out that our experimental
results in the two cases are both close to 30\% of the
theoretically allowed maximum value.

During the above calculation, we
have used the worst-case results in every step for the security.
Obviously, there are more economic methods for the calculation of
final key rate. Here we have not considered the consumption of raw
keys for QBER test. Now we reconsider the key rate calculation of decoy states
above. We assumed the worst case of
$s_0=(1+r_0)S_0$ and $s_0=(1-r_0)S_0$ for calculating $\Delta_1^{\mu}$
and $E_{1}^{\mu}$ respectively. Although we
don't exactly know the true value of $s_0$, there must be one fixed
value for both calculations.
Therefore we can choose every possible value in the range of
$(1-r_0)S_0\leq s_0\leq (1+r_0)S_0$ and use it to calculate
$\Delta_1^{\mu}$, $E_{1}^{\mu}$ and the final key rate,
and then pick out the smallest value as
the lower bound of decoy states key rate. Figure 3 demonstrates the
results with a larger range of $|r_0|$ ($r_0$ can be negative here) than the actual range
in the case of 13.448~km.
This economic calculation method can obtain a more tightened
value of the lower bound, which is larger than the
result using the simple calculation method above with
two-step worst-case assumption for $s_0$ values.

We have tested the system with different fiber lengths
and compared the final key rates with the theoretical
values, see Fig.2(b). The differences between them are mainly
due to the imperfect polarization compensation and the
possible statistical fluctuation.

\begin{figure}
\centerline{\includegraphics[scale=0.45]{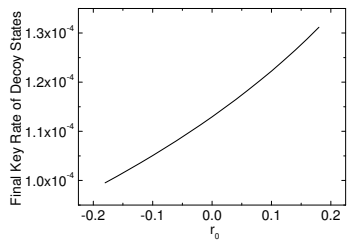}}
\caption{\label{Fig3}The final key rate of decoy pulses varies with
the vacuum counting rate of $s_0$ in the case of 13.448~km, where
$s_0=(1+r_0)S_0$.}
\end{figure}


The two measurement schemes have their own advantages.
One-detector scheme can overcome the security loopholes
of the detector efficiency mismatch and generate the
unconditionally secure final keys while the other can implement
longer distance. If we use four-detector scheme with four high quality SPDs
the final key rate and maximum distance will be improved.
Also, the balance between the efficiency and the dark counts of SPD is
important. During the experiment, we have even reduced the
efficiency of SPD purposely to reduce the dark counts to
obtain better balance. Hopefully, a low-noise and high-efficiency
detector at telecommunication wavelengths can be used in the
future to further improve the final key rate. The superconducting
transition-edge sensor is one of the promising candidates~\cite{TES}.


In summary, we implement the polarization-based one-way
decoy-state QKD over 102~km and also implement 75~km
one-way decoy-state QKD using only one detector to
really offer the unconditionally secure final keys.


This work is supported by the NNSF of China, the CAS and the
National Fundamental Research Program.



\end{document}